\documentclass[pre,eqsecnum,twocolumn]{revtex4}
\usepackage[centertags]{amsmath}
\usepackage{amssymb}
\usepackage{amsthm}

\begin{document}

\title{Inhomogeneous quantum diffusion and decay of a meta-stable state}
\author{Pulak Kumar Ghosh, Debashis Barik and Deb Shankar Ray{\footnote{
Email address: pcdsr@mahendra.iacs.res.in}}} \affiliation{Indian
Association for the Cultivation of Science, Jadavpur, Kolkata 700
032, India}

\begin{abstract}
We consider the quantum stochastic dynamics of a system whose
interaction with the reservoir is considered to be linear in bath
co-ordinates but nonlinear in system co-ordinates. The role of the
space-dependent friction and diffusion has been examined in the
decay rate of a particle from a meta-stable well. We show how the
decay rate can be hindered by inhomogeneous dissipation due
nonlinear system-bath coupling strength.
\end{abstract}

\maketitle

\section{Introduction}
The system-reservoir model\cite{zwa,cald,cald1,grab,tani,db1} has
been the standard paradigm for description  of quantum
dissipation\cite{wei} over the last few decades. Its application is
ubiquitous in physics and chemistry. In the overwhelming majority of
the situations the interaction between the system and reservoir is
considered to be linear in bath co-ordinates and linear in system
co-ordinates. This in turn relates the additive noise of the thermal
bath with linear dissipation of the system through
fluctuation-dissipation relation. However when coupling between the
reservoir is linear in bath co-ordinates but nonlinear in system
coordinates, one encounters multiplicative noise
\cite{san,san1,jay,san2,mas,lin1,sak,tai,san3,barik} and nonlinear
dissipation in the form of co-ordinate dependent friction. The role
of the space-dependent friction in classical context has been
explored in several issues, e. g., charge transfer reaction in a
polar medium and activated rate processes in the overdamped
regime\cite{mill,maka,stra,bao,pk1}, fluctuation-induced
transport\cite{but,rei}, stochastic resonance\cite{ben,ff3},
noise-induced transition\cite{tran1,d41}, etc. to name a few.

In the present paper we are concerned with space-dependent friction
and multiplicative noise in a quantum mechanical context. A standard
treatment of quantum dissipation based on linear interaction between
 the system and the reservoir was put forward in early eighties by
Caldeira and Leggett \cite{cald,cald1}, which found extensions
 and wide applications
 in several areas of condensed matter and chemical
 physics\cite{cald,cald1,wei,ff2},particularly in
 quantum interference devices,
 polaron in a magnetic field, quantum decay processes etc. However, the
 exploration of physics of nonlinear system-reservoir coupling in a
 quantum system has been the subject of recent interest only. It has
 been observed that quantum dissipation can reduce the appearance of
 meta-stable state and barrier drift in a double-well potential\cite{bao}.
 Nonlinear coupling has been modeled by Tanimura and co-workers
 \cite{tani} in
 the study of elastic and inelastic relaxation mechanisms, and their
 interplay in the Raman and infrared spectra. A quantum transport as
 a consequence state-dependent diffusion has been investigated\cite{barik} in
 a periodic potential. A closer look into the nature of nonlinear coupling
 (say, a linear in bath co-ordinate but quadratic in system
 co-ordinate) in contrast to linear coupling suggests that while the
 latter induces displacement (Fig.1(a)), the former coupling induces
 fluctuations of the system potential (Fig.1(b)). The effect of this fluctuation
 has been
 addressed in both spin-spin and spin-lattice relaxation
processes\cite{muka,oxt,bader,polla}.
 The fluctuation in
 the system potential due to nonlinear coupling, in general, tends
  to make the effective potential nonlocal
  in addition to making the diffusive process inhomogeneous
 in nature. This consideration is therefore of significant relevance
 in the context of quantum rate processes and it is worthwhile to
 examine the nature of the decay of a meta-stable state as a result of
 inhomogeneous quantum diffusion.
 Based on a coherent state
 representation of the noise operator and canonical thermal
 Wigner distribution\cite{hil} of the bath oscillators
  nonlinearly coupled to the
 system we have recently developed the quantum stochastic dynamics in
 the overdamped limit\cite{barik}. This has been carried out after adiabatic
 elimination of the momentum in the relevant Langevin dynamics which
 takes care of systematic quantum corrections to the potential.

 Our aim in this paper is to
 search for the quantum signature of this nonlinear coupling on the
 decay rate, particularly in the pre-exponential factor and
the generalized effective potential.
 We have emphasized how to take care of correct order of
 dissipation and appropriate thermodynamic consistency to avoid
 spurious term in the steady state distribution\cite{san1,san2} and
 show how the decay rate can be reduced by inhomogeneous
 dissipation.

 The outlay of the paper is as follows: In Sec.II we introduce
 briefly the basic aspects of quantum stochastic dynamics with
 nonlinear coupling on a general setting. This is followed
  by a quantum Langevin equation in the overdamped
 limit.
  Sec.III is devoted
  to the calculation of decay
 rate from a meta-stable state under the influence of inhomogeneous
 diffusion and dissipation. This paper is concluded in Sec.IV.

\section{Quantum stochastic dynamics with non-linear coupling}

We consider a particle of unit mass coupled to a medium comprised of
a set of harmonic oscillators with frequency $\omega_j$. This is
described by the following system-bath Hamiltonian \cite{bao,tani}.

\begin{equation}\label{2.1}
\hat{H}=\frac{\hat{p}^2}{2}+V(\hat{q})+\sum_j
\left[\frac{\hat{p}^2_j}{2}+ \frac{1}{2}\left(\omega_j \hat{x}_j -
\frac{c_j}{\omega_j} f(\hat{q})\right)^2\right]
\end{equation}

Here $\hat{q}$ and $\hat{p}$ are the coordinate and momentum
operators of the particle and the $\{\hat{x}_j,\hat{p}_j\}$ are the
set of coordinate and momentum operators for the bath oscillators
with unit mass. The system particle is coupled to the bath
oscillators nonlinearly through the general coupling terms
$\frac{c_j}{\omega_j}f(\hat{q})$. $c_j$ is the coupling strength.
The classical counterpart \cite{lin} of the form (\ref{2.1}) is
known for many years and also the nonlinear coupling in quantum
system has been studied in a few occasions \cite{bao,tani}. The
potential $V(\hat{q})$ is due to external force field for the system
particle. The coordinate and momentum operators follow the usual
commutation relations $[\hat{q}, \hat{p}]=i \hbar$ and $[\hat{x}_j,
\hat{p}_k]=i \hbar \delta_{jk}$.

Eliminating the bath degrees of freedom in the usual way we obtain
the operator Langevin equation\cite{db1,barik} for the particle

\begin{eqnarray}
\dot{\hat{q}}(t)&=&\hat{p}(t)\label{2.2}\\
\dot{\hat{p}}(t)&=&-V^\prime (\hat{q}(t))-f^\prime
(\hat{q}(t))\int_0^t f^\prime (\hat{q}(t^\prime))
\gamma(t-t^\prime)\hat{p}(t^\prime) dt^\prime \nonumber
\\
&&\;\;\;\;\;\;\;\;\;\;\;\;\;\;+f^\prime(\hat{q}(t))\hat{\eta}(t)\label{2.3}
\end{eqnarray}
where the noise operator $\hat{\eta}(t)$ and the memory kernel
$\gamma(t)$ are given by
\begin{eqnarray}\label{2.4}
\hat{\eta} (t) &=& \sum_j  \left \{
\frac{\omega_j^2}{c_j}\;\hat{x}_j (0) - f(\hat{q}(0)) \right \}
\frac{c_j^2}{\omega_j^2} \cos \omega_j t \nonumber\\
&&+\sum_j \frac{c_j}{\omega_j} \hat{p}_j (0) \sin \omega_j t
\end{eqnarray}
and
\begin{equation}\label{2.5}
\gamma(t)=\sum_j \frac{c_j^2}{\omega_j^2} \;\cos\omega_j t
\end{equation}

It is clear from the operator Langevin equation Eq.(\ref{2.3}) for
the system that the noise operator is multiplicative and the
dissipative term is nonlinear with respect to system coordinate due
to the nonlinear coupling term in the system-bath Hamiltonian. In
the case of linear coupling, \textit{i.e.}, $f(\hat{q})=\hat{q}$
Eq.(\ref{2.3}) reduces to a quantum generalized Langevin equation
\cite{wei} in which the noise term is additive and the dissipative
term is linear.

In the Markovian limit the generalized quantum Langevin equation
Eq.(\ref{2.3}) reduces to the form

\begin{subequations}
\begin{eqnarray}
\dot{\hat{q}}(t)&=&\hat{p}(t)\label{2.6a}\\
\dot{\hat{p}}(t)&=&-V^\prime(\hat{q}(t))-\Gamma\;
[f^\prime(\hat{q}(t))]^2\; \hat{p}(t)\nonumber
\\
&&\;\;\;\;\;\;+f^\prime(\hat{q}(t)) \hat{\eta}(t)\label{2.6b}
\end{eqnarray}
\end{subequations}
where $\Gamma$ is dissipation constant in the Markovian limit.

Following Ref.\cite{db1} we then carry out a quantum mechanical
average $\langle..\rangle$ over the product separable bath modes
with coherent states and system mode with an arbitrary state at
$t=0$ in Eq.(\ref{2.6a}) and Eq.(\ref{2.6b}) to obtain a generalized
Langevin equation as
\begin{subequations}
\begin{eqnarray}
\dot{q}&=&p\label{2.7a}\\
\dot{p}&=&- V^\prime(q) +Q_V-\Gamma [f^\prime(q)]^2 p+Q_1 \nonumber
\\
&&+ f^\prime(q) \eta(t)+Q_2\label{2.7b}
\end{eqnarray}
\end{subequations}
where the quantum mechanical mean value of position and momentum
operators $\langle \hat{q}\rangle=q$ and $\langle \hat{p}\rangle=p$.
$Q_V=V^\prime(q)-\langle V^\prime(\hat{q})\rangle$ represent quantum
correction due to nonlinearity of the system potential. $Q_1=\Gamma
\left[[f^\prime(q)]^2 p-\langle [f^\prime(\hat{q})]^2
\hat{p}\rangle\right]$ and $Q_2=\eta(t)\left[\langle
f^\prime(\hat{q}) \rangle-f^\prime(q)\right]$ are also quantum
correction terms due to nonlinearity of the system-bath coupling
function.

 Furthermore quantum mechanical mean Langevin force is
given by,
\begin{eqnarray}\label{2.8}
\eta (t) &=& \sum_j  \left \{
\frac{\omega_j^2}{c_j}\;\langle\hat{x}_j (0)\rangle - \langle
f(\hat{q}(0))\rangle \right \} \frac{c_j^2}{\omega_j^2} \cos
\omega_j t \nonumber \\
&&+\sum_j  \frac{c_j}{\omega_j}\; \langle \hat{p}_j (0) \rangle \sin
\omega_j t
\end{eqnarray}
To realize $\eta(t)$ as an effective c-number noise we now introduce
the ansatz \cite{hil,skb,db1,db2,bk1,bk2} that the momentum $\langle
\hat{p}_j (0) \rangle$ and the shifted coordinates
$\{\frac{\omega_j^2}{c_j}\;\langle\hat{x}_j (0)\rangle - \langle
f(\hat{q}(0))\rangle \}$ of the bath oscillators are distributed
according to a Wigner canonical thermal distribution of Gaussian
form as:
\begin{eqnarray}
&\mathcal{P}_j \left(\{\frac{\omega_j^2}{c_j}\;\langle\hat{x}_j
(0)\rangle - \langle f(\hat{q}(0))\rangle \},\langle \hat{p}_j
(0) \rangle\right)\nonumber\\
&=\mathcal{N} \exp\left\{-\;\frac{[\;\langle \hat{p}_j
(0)\rangle^2+\frac{c_j^2}{\omega_j^2}\{\frac{\omega_j^2}{c_j}\;\langle\hat{x}_j
(0)\rangle - \langle f(\hat{q}(0))\rangle \}^2]}{2 \hbar \omega_j
\left(\bar{n}_j(\omega_j)+\frac{1}{2}\right)
}\right\}\nonumber\\
\label{2.9}
\end{eqnarray}
so that for any quantum mechanical mean value,
$\mathcal{O}_j\left(\{\frac{\omega_j^2}{c_j}\;\langle\hat{x}_j
(0)\rangle - \langle f(\hat{q}(0))\rangle \},\langle \hat{p}_j
(0)\rangle\right)$ which is a function of mean value of the bath
operators $\langle \hat{x}_j (0) \rangle$ and $\langle \hat{p}_j (0)
\rangle$, its statistical average $\langle...\rangle_S$ is

\begin{equation}\label{2.10}
\langle \mathcal{O}_j\rangle_S=\int \mathcal{O}_j \mathcal{P}_j\;
d\langle \hat{p}_j (0)\rangle\;
d\{\frac{\omega_j^2}{c_j}\langle\hat{x}_j (0)\rangle - \langle
f(\hat{q}(0))\rangle \}
\end{equation}

Here $\bar{n}_j(\omega_j)$ indicates the average thermal photon
number of the j-th oscillator at the temperature $T$ and
$\bar{n}_j(\omega_j)=[\exp(\frac{\hbar \omega_j}{ k_B T})-1]^{-1}$
and $\mathcal{N}$ is the normalization constant.

The distribution $\mathcal{P}_j$ (Eq.(\ref{2.9})) and the definition
of statistical average Eq.(\ref{2.10}) imply that c-number noise
$\eta(t)$ must satisfy
\begin{equation}\label{2.11}
\langle \eta(t) \rangle_S=0
\end{equation}
\begin{equation}\label{2.12}
\langle \eta(t) \eta(t^\prime) \rangle_S=\frac{1}{2}\sum_j
\frac{c_j^2}{\omega_j^2}\; \hbar \omega_j \left(\coth \frac{\hbar
\omega_j}{2 k_B T}\right) \cos \omega_j (t-t^\prime)
\end{equation}
In the Markovian limit the noise correlation becomes\cite{pk1,lui}
\begin{subequations}
\begin{eqnarray}
\langle \eta(t) \eta(t^\prime) \rangle_S&=&2D_0\delta(t-t^\prime)\label{2.13a}\\
D_0&=&\frac{1}{2}\Gamma \hbar \omega_0
\left(\bar{n}(\omega_0)+\frac{1}{2}\right)\label{2.13b}
\end{eqnarray}
\end{subequations}
where $\omega_0$ is the average bath frequency and the spectral
density function is considered in the Ohmic limit. The use of
Gaussian form [Eq.(\ref{2.9})] for a bath distribution needs some
elaboration. It follows from a simple calculation of partition
function of a nonlinearly coupled harmonic oscillators, that a
Gaussian form of ansatz is not valid, in general. In the present
case we emphasize that one is concerned here with a system-reservoir
nonlinear coupling where the coupling is nonlinear in system
coordinates but still linear in bath coordinates, so that the use of
coherent state basis for harmonic oscillators enable us to factor
out the dependence of nonlinear system coordinate in our
calculation. This is also reflected in the fluctuation-dissipation
relation (\ref{2.12}) which retains its usual form, in spite of
nonlinear system-reservoir interaction. Further to this we emphasize
that the overdamped situation in our treatment refers to large
$\Gamma$ rather than $\Gamma\;[f'(x)]^2$. For a better validity of
the theory the nonlinearity should not be too strong.

The equations (\ref{2.11}) and (\ref{2.12}) imply that the c-number
noise $\eta(t)$ is such that it is zero centered and satisfies the
standard fluctuation-dissipation relation. We thus identify
$\eta(t)$ as a classical looking noise with quantum mechanical
content. The quantum Langevin equation can now be rewritten as
follows:
\begin{equation}\label{2.14}
\dot{q}= p
\end{equation}
\begin{eqnarray}
\dot{p}&=& -V^\prime(q)+Q_V-\Gamma [f^\prime(q)]^2p- 2 \Gamma\; p\;
f^\prime(q) Q_f -\Gamma \;p\; Q_3 \nonumber\\ &-&2\Gamma f^\prime(q)
Q_4 -\Gamma \;Q_5 + f^\prime(q)\;\eta(t) +Q_f \;\eta(t)\label{2.15}
\end{eqnarray}
The above equation is characterized by a classical force term,
$V^\prime$, as well as its correction $Q_V$. The terms containing
$\Gamma$ are nonlinear dissipative terms where $Q_f$, $Q_3$, $Q_4$
and $Q_5$ are due to associated quantum contribution in addition to
classical nonlinear dissipative term $\Gamma [f^\prime(q)]^2 p$. The
explicit expressions for $Q_V$,  $Q_f$, $Q_3$, $Q_4$ and $Q_5$ are
given in the Appendix-A. The last term in the above equation
contains a quantum multiplicative noise term in addition to the
usual classical contribution $f^\prime(q) \eta(t)$. The classical
limit of the above equation was derived earlier by Lindenberg and
Seshadri \cite{lin}. Furthermore quantum dispersions due to
potential and coupling terms in the Hamiltonian are entangled with
nonlinearity. The quantum noise due heat bath on the other hand is
expressed in terms of the fluctuation-dissipation relation.

 While considering hydrodynamic interaction, \textit{i.e.},
when the fluctuation is position/state dependent or equivalently
when the noise is multiplicative with respect to system variables
the conventional adiabatic reduction of fast variables does not work
correctly. To obtain a correct equilibrium distribution Sancho
\textit{et al} \cite{san1} had proposed an alternative approach to
Langevin equation in the case of multiplicative noise system. By
carrying out a systematic expansion of the relevant variables in
powers of $\Gamma^{-1}$ and neglecting terms smaller than
$O(\Gamma^{-1})$ they obtained a Langevin equation corresponding to
a Fokker-Planck equation in position space. This description leads
to the correct stationary probability distribution of the system
with coordinate-dependent friction.

We follow the same procedure in the quantum mechanical context. In
this limit the transient correction terms $Q_4$ and $Q_5$ do not
affect the dynamics of the position which varies in a much slower
time scale. So the quantum Langevin equation Eq.(\ref{2.14}) and
Eq.(\ref{2.15}) can be written as, respectively,
\begin{eqnarray}
\dot{q}&=&p\label{2.16}\\
\dot{p}&=&-V^\prime(q)+Q_V-\Gamma\; h(q) p+g(q)\;
\eta(t)\label{2.17}
\end{eqnarray}
where
\begin{eqnarray}
h(q) &=& [f^\prime(q)]^2+2 f^\prime(q) Q_f +Q_3\label{2.18}\\
g(q) &=& f^\prime(q)+Q_f\label{2.19}
\end{eqnarray}
The variable $g(q)$ arises due to nonlinearity of the system-bath
coupling function $f(q)$, $Q_f$ being the quantum correction to the
classical contribution $f'(q)$. For a linear coupling function
$g(q)$ reduces to a constant.

 Following the method of Sancho \textit{et al} \cite{san1}
further we obtain the Fokker-Planck equation in position space
corresponding to Langevin equation Eq.(\ref{2.17})

\begin{eqnarray}
\frac{\partial P(q,t)}{\partial t}&=&\frac{\partial}{\partial
q}\left[ \frac{V^\prime(q)-Q_V}{\Gamma\; h(q)} \right]
P(q,t)\nonumber\\
&+&D_0\frac{\partial}{\partial q}\left[\frac{1}{\Gamma\;(h(q))^2}\;
g^\prime(q) g(q) \right] P(q,t)\nonumber\\
&+& D_0\frac{\partial}{\partial
q}\left[\frac{g(q)}{\Gamma\;h(q)}\;\frac{\partial}{\partial
q}\;\frac{g(q)}{\Gamma\;h(q)}\right]P(q,t)\nonumber\\
\label{2.20}
\end{eqnarray}

 The Stratonovich prescription leads to corresponding
 Langevin equation as given by

\begin{equation}\label{2.21}
\dot{q}=-\;\frac{V^\prime(q)-Q_V}{\Gamma\;h(q)}-D_0\;\frac{g^\prime(q)
g(q)}{\Gamma\;(h(q))^2}\; +\frac{g(q)}{\Gamma\;h(q)}\;\eta(t)
\end{equation}

Eq.(\ref{2.21}) is quantum Langevin equation for multiplicative
noise with position dependent friction in the overdamped limit
(\textit{i.e.}, corrected upto $O(1/\Gamma)$).

In the classical limit, \textit{i.e.}, $\hbar\omega_0 \ll k_B T$,
$h(q)=[f^\prime(q)]^2$, $g(q)=f^\prime(q)$, $Q_V=0$ and
$D_0=\Gamma\;k_B T$ so the quantum Langevin equation reduces to its
classical counterpart as derived by Sancho \textit{et al}
\cite{san1}
\begin{equation}\label{2.22}
\dot{q}=\frac{1}{\Gamma\;[f^\prime(q)]^2}
\left[-V^\prime(q)-\Gamma\;k_B T\;
\frac{f^{\prime\prime}(q)}{f^\prime(q)}+f^\prime(q)\;\eta(t) \right]
\end{equation}

\section{Decay rate from a metastable state}

Attention is now restricted here to the problem of barrier crossing
dynamics in the presence of nonlinear system-bath coupling. The
particle coordinate $q$ (which in our case is the quantum mechanical
mean position) corresponds to the reaction coordinate and its values
at the minima of $V(q)$ denote the reactant and product states
separated by a finite barrier, the top being a metastable state
representing the transition state. We choose the potential and the
coupling function of the following forms ;
\begin{eqnarray}
V(q) &=& \frac{1}{2}a\;q^2-\frac{1}{3}b\;q^3\label{3.1}\\
f(q) &=& b_1\;q+\frac{1}{2}b_2\;q^2\label{3.2}
\end{eqnarray}
$a$ and $b$ are the potential parameters and the maximum and the
minimum of the potential are at the point $q=q_b=a/b$ and $q=q_a=0$,
  respectively. To clarify the role of the nonlinear system-bath
coupling we now turn to Eq.(\ref{2.1}), in which system-bath
interaction is given by ($\omega_j$ is absorbed in $c_j$),
\begin{eqnarray}
\hat{H}_{int}&=&-\sum c_j\;\hat x_j f(\hat{q})\nonumber
\\
&=&-(c_1\hat x_1+c_2\hat x_2+...)(b_1\hat
q+\frac{1}{2}b_2\;\hat{q}^2)\label{3.3}
\end{eqnarray}
The linear-linear coupling term ($\sum c_j \;b_1\hat q$) is
proportional to $b_1$ while square-linear coupling term ($\sum c_j
\;b_2\hat q^2$) is proportional to $b_2$. The linear-linear coupling
model a very commonly used mechanism for studying quantum
dissipative dynamics, e. g., energy dissipation from vibrational
system mode to the heat bath modes during population decay. In
addition to energy dissipation due to linear-linear coupling, one
also encounters fluctuation in the system potential due to nonlinear
coupling. We refer to Fig.1(b) where a schematic illustration of the
fluctuation of system-bath coupling in a relevant potential well is
presented. Since this fluctuation is the key mechanism for
state-dependent diffusion, it is apparent that the calculation of
transition rate in presence of state-dependent diffusion differs
from the calculation of escape rate for particles subject to
thermally uniform noise, in that we may conveniently substitute the
role of potential $V(q)$ by a generalized potential $\Psi (q)$
corresponding to the former case. A simple way to determine
transition rate is to consider a steady state condition. The
Fokker-Planck equation Eq.(\ref{2.20}) in the overdamped limit can
be rewritten in a more compact form as
\begin{equation}\label{3.4}
\frac{\partial P(q,t)}{\partial t}=\frac{\partial}{\partial
q}\;\frac{1}{\Gamma\;h(q)}\left[V^\prime(q)-Q_V+\frac{D_0}{\Gamma}\;\frac{\partial}{\partial
q }\;\frac{g(q)^2}{h(q)}\right]P(q,t)
\end{equation}
In the overdamped limit the stationary current from Eq.(\ref {3.4})
is given by
\begin{equation}\label{3.5}
J=-\;\frac{1}{\Gamma\;h(q)}\left[V^\prime(q)-Q_V+\frac{D_0}{\Gamma}\;\frac{d}{dq}
\left(\frac{g(q)^2}{h(q)}\right)\right]P_{st}(q)
\end{equation}
The Eq.(\ref{3.5}) can be rearranged into the following form,
\begin{equation}\label{3.6}
\frac{d}{dq}\left\{P_{st} \;L(q)\;\exp\left[\frac{ \Psi
(q)}{D_q}\right] \right\}=-\frac{J\;\Gamma
}{D_q}\;h(q)\exp\left[\frac{ \Psi (q)}{D_q}\right]
\end{equation}
where
\begin{eqnarray}\label{3.7}
\Psi(q)= \int^q
\frac{V'(q')-Q_V}{L(q')}\;dq';\;\;\;\;\;\;\;\;L(q)=\frac{g(q)^2}{h(q)}
\end{eqnarray}
and  $D_q=\frac{D_0}{\Gamma}$. To calculate a steady state solution
we assume an absorbing boundary
 at a coordinate $q_s$ past
the intervening potential maximum at $q=q_b$. So integrating
Eq.(\ref{3.6}) between $q=q_a$ and $q=q_s$, we have
\begin{equation}\label{3.8}
J= \frac{P_{st}(q_a)\;L(q_a)\;\exp\left[\frac{ \Psi
(q_a)}{D_q}\right]}{\int_{q_a}^{q_s}\frac{\Gamma
}{D_q}\;h(q)\exp\left[\frac{ \Psi (q)}{D_q}\right]\;dq}
\end{equation}
The steady escape rate over the top of the potential is given by
flux over population,
\begin{equation}\label{3.9}
k=\frac{J}{n_a}
\end{equation}
where $n_a$ is the population around the potential minimum ($q_a$).
The zero current defines an equilibrium condition in the
neighborhood of $q_a$. Thus with $J=0$, probability distribution
function near the potential minimum is given by
\begin{equation}\label{3.10}
P_{st}(q)=P_{st}(q_a)\frac{L(q_a)}{L(q)}\exp\left[-\int_{q_a}^q
dq'\;\frac{V'(q')-Q_V}{L(q')}\;\frac{\Gamma}{D_0}\right]
\end{equation}
In terms of the generalized potential Eq.(\ref{3.10}) can be
rewritten as
\begin{equation}\label{4.11}
P_{st}(q)=P_{st}(q_a)\frac{L(q_a)}{L(q)}\exp\left[\frac{ -\Psi
(q)}{D_q}+\frac{ \Psi (q_a)}{D_q}\right]
\end{equation}
Population around the bottom of the potential is given by
\begin{eqnarray}\label{3.12}
n_a&=&\int_{q_1}^{q_2}P_{st}(q)\;dq\nonumber
\\
&=&P_{st}(q_a)\;L(q_a)\; \exp\left[\frac{ \Psi (q_a)}{D_q}\right]
\int_{q_1}^{q_2}\frac{1}{L(q)}\exp\left[\frac{ -\Psi
(q)}{D_q}\right]\nonumber
\\
\end{eqnarray}
where $q_1$ and $q_2$ are two points around potential minimum. So
the expression for transition rate is given by
\begin{equation}\label{3.13}
k = \frac{ D_q}{\Gamma}
\;\frac{1}{\int_{q_s}^{q_a}dq\;h(q)\exp\left[\frac{ \Psi
(q)}{D_q}\right]}\;\frac{1}{\int_{q_1}^{q_2}dq
\;\frac{1}{L(q)}\exp\left[\frac{ -\Psi (q)}{D_q}\right]}
\end{equation}
By linearizing the generalized potential around maximum and minimum
and evaluating the integrals in the usual way we obtain
\begin{equation}\label{3.14}
k=\frac{|\Psi''(q_a)|^{\frac{1}{2}}\;|\Psi''(q_b)|^{\frac{1}{2}}L(q_a)}{2\pi
\Gamma h(q_b)}\exp\left[-\left\{\frac{ \Psi (q_b)- \Psi (q_a)}{D_q}
\right\}\right]
\end{equation}
 $\Psi(q)$ can be simplified further to yield
\begin{eqnarray}\label{3.15}
\Psi(q)&=& \int^q \frac{V'(q')-Q_V}{L(q')}\;dq'\nonumber \\
&=&\int^q dq'
\left(V'(q')-Q_V\right)\left(1+\frac{Q_3-Q_f^2}{(f'(q')+Q_f)^2}\right)\nonumber
\\
\end{eqnarray}
The rate constant (\ref{3.14}) then reduces to the following
expression,
\begin{eqnarray}\label{3.16}
k&=&\frac{\sqrt{[\omega_a^2(1+\Delta_1)-\Delta_{v_1}][\omega_b^2(1+\Delta_2)-
\Delta_{v_2}]}}{2\pi\Gamma \;h(q_b)\;(1+\Delta_1)}\nonumber\\
&& \times \exp\left[-\left\{\frac{ \Psi (q_b)- \Psi (q_a)}{D_q}
\right\}\right]
\end{eqnarray}
where $\omega_a=V''(q_a)$ and $\omega_b=V''(q_b)$ are the
frequencies at the bottom and barrier top, respectively. $\Delta_1$,
$\Delta_2$, $\Delta_{v_1}$ and $\Delta_{v_2}$ are the contributions
due to quantum correction and are given by,
\begin{eqnarray}\label{3.17}
\Delta_1=\left[\frac{Q_3-Q_f^2}{(f'(q)+Q_f)^2}\right]_{q=q_a}
\Delta_2=\left[\frac{Q_3-Q_f^2}{(f'(q)+Q_f)^2}\right]_{q=q_b}\nonumber
\\
\end{eqnarray}
\begin{eqnarray}\label{3.18}
\Delta_{v_1}&=&\left[\frac{d}{dq}\left\{
Q_V\left(1+\frac{Q_3-Q_f^2}{(f'(q)+Q_f)^2}\right)\right\}\right]_{q=q_a}\nonumber
\\
\Delta_{v_2}&=&\left[\frac{d}{dq}\left\{
Q_V\left(1+\frac{Q_3-Q_f^2}{(f'(q)+Q_f)^2}\right)\right\}\right]_{q=q_b}
\end{eqnarray}
 The quantum nature of the
transition rate (\ref{3.16}) is manifested through the quantum
correction due to nonlinearity of the potential and coupling
function and the quantum inhomogeneous diffusion. It also contains
the signature of nonlinear coupling in the pre-exponential factor
and in the generalized potential $\Psi(q)$. More generally, the
effect of state-dependent diffusion makes its presence felt in the
dynamics of decay.

In order to check some inherent consistency of the rate expression
we now examine the following situations.

(a) First consider that the coupling function is linear, i. e.,
$b_2=0$ and $f(q)=q$ ($b_1$ is assumed to be unity). Then the
generalized potential is given by
\begin{eqnarray}\label{3.19}
\Psi(q)=\int^q dq (V'(q')-Q_V)
\end{eqnarray}
and the transition rate reduces to
\begin{eqnarray}\label{3.20}
k&=&\frac{\sqrt{[\omega_a^2-\Delta_{v_1}][\omega_b^2-
\Delta_{v_2}]}}{2\pi\Gamma }\exp\left[-\left\{\frac{ \Delta
\phi}{D_q} \right\}\right]\nonumber
\\
&&\times
 \exp\left[-\left\{\frac{ V (q_b)- V
(q_a)}{D_q} \right\}\right]
\end{eqnarray}
where
\begin{eqnarray}\label{3.21}
\Delta \phi=\int_{q_a}^{q_b}\frac{Q_V(q)}{D_q}\;dq
\end{eqnarray}
This quantify the characteristic quantum decay rate from a
meta-stable state for particles subject to thermally uniform noise.
On the other hand as the classical limit is approached all the
quantum correction terms ($\Delta_{v_1},\;\Delta_{v_2}$ and $\Delta
\phi$) due to nonlinearity of the potential tend to vanish and the
quantum coefficient($D_q$) gets replaced by the
 classical one ($D_q \rightarrow k_b T$).
The transition rate (\ref{3.20}) reduces exactly to Kramers rate in
the overdamped limit.
\begin{eqnarray}\label{3.22}
k=\frac{\omega_a\omega_b}{2\pi\Gamma }
 \exp\left[-\left\{\frac{ V (q_b)- V
(q_a)}{k_bT} \right\}\right]
\end{eqnarray}

(b) In order to check the thermodynamic consistency we consider
further a symmetric periodic potential with periodicity $2\pi$,
\textit{i.e.}, $V(q)=V(q+2\pi)$ and periodic derivative of coupling
function with the same periodicity as that of the potential,
\textit{i.e.}, $f^\prime(q)=f^\prime(q+2\pi)$. Such a potential had
been subject of interest in the problem of classical\cite{but} and
quantum transport\cite{barik} in inhomogeneous media.

Since the potential is periodic, $Q_V$ is also a periodic function
as $Q_V=V(q)-\langle V(\hat{q})\rangle$. Similarly $Q_f$ is also
periodic as $Q_f=\langle f^\prime(\hat{q})\rangle -f^\prime(q)$, and
also $Q_3+2f^\prime(q)Q_f$ is also periodic since
$Q_3+2f^\prime(q)Q_f=\langle
[f^\prime(\hat{q})]^2\rangle-[f^\prime(q)]^2$. So from
Eqs.(\ref{2.18}) and (\ref{2.19}) it is clear that $h(q)$ and $g(q)$
are also periodic functions of $q$ with periodicity $2\pi$. Now it
is easy to check that $\Psi(q)$ is also a periodic function with
periodicity $2\pi$.
\begin{eqnarray}\label{3.23}
\Psi(q)=\Psi(q+2\pi)
\end{eqnarray}
Consider the bottom of the potential well at $q_a=2\pi n$ and
potential maxima at $q_b=(2n+1)\pi$. Then with the help of
Eq.(\ref{3.16}) we find a transition rate from valley $n$ to valley
$n+1$
\begin{eqnarray}\label{3.24}
k_{n+1,n}&=&\frac{\sqrt{[\omega_a^2(1+\Delta_1)-\Delta_{v_1}][\omega_b^2(1+\Delta_2)-
\Delta_{v_2}]}}{2\pi\Gamma \;h(q_b)\;(1+\Delta_1)}\nonumber
\\
&&\times\exp\left[-\left\{\frac{ \Psi (q_b)- \Psi (q_a)}{D_q}
\right\}\right]
\end{eqnarray}
Repeating these considerations for the transition rate from valley
$n+1$ to valley $n$ yields
\begin{eqnarray}\label{3.25}
k_{n,n+1}&=&\frac{\sqrt{[\omega_a^2(1+\Delta_1)-\Delta_{v_1}][\omega_b^2(1+\Delta_2)-
\Delta_{v_2}]}}{2\pi\Gamma \;h(q_b)\;(1+\Delta_1)} \nonumber
\\
&&\times\exp\left[-\left\{\frac{ \Psi (q_b)- \Psi (q_a+2\pi)}{D_q}
\right\}\right]
\end{eqnarray}
Thus the average particle velocity to a preferential direction is
given by
\begin{eqnarray}\label{3.26}
\langle\frac{dq}{dt}\rangle &=&
2\pi\left(k_{n+1,n}-k_{n,n+1}\right)\nonumber\\
 &=& 2\pi
k_{n+1,n}\left(1- \exp\left[-\left\{\frac{ \Psi (q_a)- \Psi
(q_a+2\pi)}{D_q} \right\}\right] \right)\nonumber
\\
&=&0
\end{eqnarray}
We thus conclude that there is no occurrence of current for a
periodic potential and periodic derivative of coupling with same
periodicity. At the macroscopic level this confirms that there is no
generation of perpetual motion of second kind, \textit{i.e.}, no
violation of second law of thermodynamics. Therefore the
thermodynamic consistency based on symmetry considerations ensures
the validity of the present formalism.

Before we proceed to numerical results it is important to clarify
the issues regarding the calculation of quantum correction terms.
The details of the calculations of quantum correction terms are
shown in the Appendix-A. One can calculate the  value of quantum
dispersion terms $\langle\delta\hat{q}^n\rangle$ by direct numerical
simulation of the coupled Eqs. (A10) subject to appropriate boundary
conditions. It is also instructive to deal with the quantum
correction terms in the analytical way to find out the approximate
value of quantum dispersion terms.
 For overdamped limit we neglect
the $\delta\dot{\hat{p}}$ term from Eq. (\ref{A9}) to obtain

\begin{eqnarray}\label{3.27}
\frac{d}{dt}\;\delta\hat{q}&=&\frac{1}{\Gamma\;[f^\prime(q)]^2}
\left[-V^{\prime\prime}(q)\;
\delta\hat{q}-2\Gamma\;p\;f^\prime(q)f^{\prime\prime}(q)\delta\hat{q}\right]\nonumber
\\
&+&\frac{1}{\Gamma\;[f^\prime(q)]^2} \left[\eta(t)
f^{\prime\prime}(q)\delta\hat{q}\right]+O(\delta\hat{q}^2)
\end{eqnarray}

 With the help of Eq. (\ref{3.27}) we then obtain the
 equations for $\langle\delta\hat{q}^n\rangle$
\begin{eqnarray}\label{3.28}
\frac{d}{dt}\langle\delta\hat{q}^2\rangle&=&\frac{2}{\Gamma\;[f^\prime(q)]^2}\left[
-V^{\prime\prime}(q)\langle\delta\hat{q}^2\rangle\right]\nonumber
\\
&-& \frac{4}{\Gamma\;[f^\prime(q)]^2}\left[\Gamma\;p\;f^\prime(q)
f^{\prime\prime}(q)\langle\delta\hat{q}^2\rangle\right]\nonumber
\\
&+& \frac{2}{\Gamma\;[f^\prime(q)]^2}\left[\eta(t)
f^{\prime\prime}(q)\langle\delta\hat{q}^2\rangle\right]+
O(\langle\delta\hat{q}^3 \rangle)\nonumber
\\
\end{eqnarray}
\begin{eqnarray}\label{3.29}
\frac{d}{dt}\langle\delta\hat{q}^3\rangle &=&
\frac{3}{\Gamma\;[f^\prime(q)]^3}\left[
-V^{\prime\prime}(q)\langle\delta\hat{q}^3\rangle\right]\nonumber
\\&-&\frac{6}{\Gamma\;[f^\prime(q)]^3}\left[\;\Gamma\;p\;f^\prime(q)
f^{\prime\prime}(q)\langle\delta\hat{q}^3\rangle\right]\nonumber
\\
&+& \frac{3}{\Gamma\;[f^\prime(q)]^3}\left[\eta(t)
f^{\prime\prime}(q)\langle\delta\hat{q}^3\rangle\right]+
O(\langle\delta\hat{q}^4 \rangle)\nonumber
\\
\end{eqnarray}
and so on. (It is apparent from Eqs. (\ref{3.28}-\ref{3.29}) that in
the overdamped limit the higher order quantum contributions are
small since each successive order of correction is lower than the
preceding one by a factor of $\frac{1}{\Gamma}$)

 A simplified expression for the leading order quantum
correction term $\langle\delta\hat{q}^2\rangle$ can be estimated by
neglecting the higher order coupling terms in the square bracket in
Eq.(\ref{3.28}) (since the nonlinearity of the potential is small
the terms of the order of $f''(x) $ are small) and rewriting it as
$d\langle\delta\hat{q}^2
\rangle=\frac{2}{\Gamma\;[f^\prime(q)]^2}\;V^{\prime\prime}(q)
\langle\delta\hat{q}^2\rangle \; dt$. The overdamped deterministic
mean motion on the other hand gives
$dq=-\;\frac{V^\prime(q)}{\Gamma\;[f^\prime(q)]^2}\;dt$. These
together yield after integration\cite{d41}
\begin{equation}\label{3.30}
\langle\delta\hat{q}^2\rangle=\Delta_q [V^\prime(q)]^2
\end{equation}
where
$\Delta_q=\frac{\langle\delta\hat{q}^2\rangle_0}{[V^\prime(q_0)]^2}$
 and $q_0$ is a quantum mechanical mean position at which
$\langle\delta\hat{q}^2\rangle$ becomes minimum, i. e.,
$\langle\delta\hat{q}^2\rangle_{q_0}=\frac{1}{2}\hbar /\omega_0$,
$\omega_0$ being defined earlier.

The relevant quantum correction terms (\ref{A1}), (\ref{A4}) and
(\ref{A5}) can be rewritten for the potential(\ref{3.1}) and
coupling term (\ref{3.2}) as follows,
\begin{eqnarray}\label{3.31}
Q_V&=&-\frac{1}{2}V'''(q)\langle\delta\hat{q}^2\rangle\;,\\
Q_f&=&0\;,\\ \label{3.32}
 Q_3&=&[f''(q)]^2\langle\delta\hat{q}^2\rangle\;.
\label{3.33}
\end{eqnarray}
$V''''(q)$ and $f'''(q)$ vanish for the chosen metastable potential
and square-linear coupling. So by calculating the quantum dispersion
term $\langle\delta\hat{q}^2\rangle$ we can find out the values of
$\Delta_1,\;\Delta_2,\;\Delta_{v_1},\;\Delta_{v_2}$ and $\Delta
\phi$ as given by Eqs(\ref{3.17}-\ref{3.18}) and (\ref{3.21}). The
corrections (\ref{3.31}) and (\ref{3.33}) are reminiscent of the
perturbation corrrections to Kramers escape rate due to
color\cite{mm1}. It may also be mentioned that important nonlinear
terms are those represented by a linear chain coupled to a linear
heat bath\cite{mm2} and in spite of the different coupling
Hamiltonian, the effects on the relevant rates are similar.

We now proceed to illustrate numerically the behavior  of quantum
transition rate given by Eq.(\ref{3.16}). It appears that in
addition to the barrier height the pre-factor is also affected by
the quantum correction terms. The calculated quantum transition rate
Eq.(\ref{3.16}) implies that both the barrier height and the
frequency factor contain the effects of
 quantum corrections due to
nonlinearity of system potential and coupling function. The effect
of quantization of the reservoir and quantum correction due to
system nonlinearity is apparent in Fig.2 in the variation of $\ln k$
with $1/T$ (Arrhenius plot) for varied contribution of nonlinear
coupling. All the curves exhibit linearity at higher temperature and
nonlinear variation is observed at lower temperature. Decrease in
$b_2$ (square-linear coupling contribution is proportional to $b_2$)
results in enhancement of transition rate. This physically implies
that with increased values of $b_2$ potential well becomes
comparatively steeper (as shown in the Fig.1) and hence the
transition rate decreases.

The effect of quantization of a classical transition rate is shown
in Fig.3, where we make a comparison of $k$ vs $1/T$ profiles for
classical and quantum cases. One observes that at low temperature
regime the classical transition rate is significantly lower in
magnitude than quantum rate and at higher temperature the effect of
quantization become insignificant. To clarify the effects of
quantization we present also an Arrhenius plot comparing the
classical and quantum regimes in the inset of Fig.3. From the inset
plot it is clear that at the high temperature regime the plot
exhibits linearity in the both cases, which is the standard
Arrhenius classical result. In low temperature limit, however, one
observes a much slower variation in the quantum case than that of
classical one. To examine this in more detail it is not difficult to
see that the exponential factor in the rate expression
Eq.(\ref{3.16}) can be reduced to a form containing the usual
Arrhenius term $\exp\left[{ -\left(
\frac{\psi(q_b)-\psi(q_a)}{kT}\right)}\right]$ times a $T^2$
enhancement term at low temperature of the form $\exp\left[{ \hbar
\omega\left( \frac{\psi(q_b)-\psi(q_a)}{2{(kT)}^2}\right)}\right]$.
The full quantum behavior can be
 interpreted in terms of an interplay between the quantum
 diffusion coefficient $D_q$ and the quantum correction due to
 nonlinearity of the system potential appearing in $\Psi(q)$.
When the temperature of the system is very low, i. e., in the vacuum
limit or in the deep tunneling region the anharmonic terms in the
potential do not contribute significantly. On the other hand as the
temperature of the system increases significantly, $D_q$ increases
resulting in decrease of the effective potential and hence $D_q$ and
$Q$ (quantum correction terms) compete to cancel the effect of each
other at higher temperature.

\section{Conclusion}
Based a theory of stochastic dynamics of a quantum particle in
inhomogeneous media under overdamped condition we have calculated
the rate of decay of a meta-stable state of the system under the
influence of quantum state dependent friction and diffusion. It has
been shown that the nonlinear interaction between the system and
bath have its imprints on the pre-exponential factor as well as on
the form of a generalized effective potential. Furthermore the
nonlinearity of the effective potential can be identified as a
typical consequence of quantum effect. The present theory can also
be extended to other selected areas of chemical physics where the
fluctuations in the potential in the system-reservoir coupling are
important. From a purely quantum mechanical point of view such
interactions give rise to multi-photon transition terms in the
dynamics. Thus their contribution to phase relaxation in addition to
population decay is of considerable significance. The difference
between linear-linear and quadratic-linear coupling in the
system-reservoir interactions terms is likely to be also important
if one calculates the time evolution of flux-flux correlation
functions\cite{mm5} which in consequence gives rise to
time-dependent rate constant. The focus of the present article being
the non-equilibrium steady state rate constant and ensuring a
correct description of equilibrium in the quantum system which often
remains vague in many related situations,we feel that an elaboration
on this issue needs more detailed consideration. We note, in
passing, that the treatment presented here concerns stochastic
processes under overdamped condition in the Markovian limit.
Extension of the theory to non-Markovian and weak friction regime is
worth-pursuing in the context of rate theory and  related contexts.

\acknowledgments Thanks are due to the Council of Scientific and
industrial research, Govt. of India, for partial financial support.

\appendix
\begin{appendix}
\section{quantum correction terms}
Referring to the quantum nature of the system in the Heisenberg
picture we now write the system operators $\hat{q}$ and $\hat{p}$ as
 $\hat{q}  =  q + \delta \hat{q}$ and $\hat{p} =  p+\delta \hat{p}$
 respectively. $\delta\hat{q}$ and $\delta\hat{p}$
 represent  quantum fluctuations around their respective mean values. By
construction $\langle \delta\hat{q}\rangle=\langle
\delta\hat{p}\rangle=0$ and they also follow the usual commutation
relation $[\delta\hat{q}, \delta\hat{p}]=i\hbar$. Using a Taylor
series expansion in $\delta\hat{q}$ around $q$ and $\delta\hat{p}$
around $p$ we express $Q_V$, $Q_1$, $Q_2$ and $Q_3$, $Q_4$, $Q_5$,
and $Q_f$ as functions of $q$, $p$, $\langle \delta\hat{q}^n\rangle$
and $\langle \delta\hat{p}^n\rangle$ as follows;
\begin{eqnarray}
Q_V&=&V^\prime(q)-\langle V^\prime(\hat{q}) \rangle\nonumber\\
& = & -\sum_{n\geq 2} \frac{1}{n!} V^{n+1}(q) \langle \delta
\hat{q}^n \rangle\label{A1}\\
Q_1&=&\Gamma \left[[f^\prime(q)]^2 p-\langle [f^\prime(\hat{q})]^2
\hat{p}\rangle\right]\nonumber\\
&=& -\Gamma\; [ 2 \;p\; f^\prime
(q)Q_f + p\; Q_3 + 2 f^\prime (q) Q_4 +Q_5]\label{A2}\\
Q_2&=& \eta(t)\left[\langle f^\prime(\hat{q})
\rangle-f^\prime(q)\right]\nonumber\\
 &=&\eta(t)\; Q_f
\label{A3}
\end{eqnarray}
where,
\begin{eqnarray}
Q_f &=& \sum_{n\geq 2}\frac{1}{n!} f^{n+1}(q) \langle \delta
\hat{q}^n \rangle\label{A4}\\
Q_3 & = & \sum_{m\geq 1}\sum_{n\geq
1}\frac{1}{m!}\frac{1}{n!}f^{m+1}(q)f^{n+1}(q) \langle \delta
\hat{q}^m \delta \hat{q}^n \rangle\label{A5}\\
Q_4 &=&\sum_{n\geq 1}\frac{1}{n!} f^{n+1}(q) \langle \delta
\hat{q}^n\delta\hat{p} \rangle\label{A6}\\
Q_5 & = & \sum_{m\geq1}\sum_{n\geq
1}\frac{1}{m!}\frac{1}{n!}f^{m+1}(q)f^{n+1}(q) \langle \delta
\hat{q}^m \delta \hat{q}^n \delta \hat{p}\rangle\nonumber
\\\label{A7}
\end{eqnarray}
The dynamics of these correction terms can be calculated\cite{barik}
with the help of the following equations, which can be derived using
the operator Langevin equations (\ref{2.6a}) and (\ref{2.6b}) and by
carrying out quantum mechanical average over the initial product
separable coherent bath states.
\begin{eqnarray}
\dot{\delta\hat{q}}&=&\delta\hat{p}\label{A8}
\end{eqnarray}
\begin{eqnarray}
&&\dot{\delta\hat{p}}=-V^{\prime\prime}(q)\delta\hat{q}-\sum_{n\ge
2}\frac{1}{n!}V^{n+1}(q)\left[\delta\hat{q}^n-\langle\delta\hat{q}^n\rangle\right]\nonumber\\
&-&2\Gamma f^\prime(q)f^{\prime\prime}(q)\delta\hat{q} +2\Gamma
f^\prime(q)\sum_{n\ge2}\frac{1}{n!}f^{n+1}(q)\left[\delta\hat{q}^n-\langle
\delta\hat{q}^n\rangle\right.\nonumber\\
&+&\sum_{m\ge 1}\sum_{n\ge
1}\frac{1}{m!}\frac{1}{n!}f^{m+1}(q)f^{n+1}(q)\left[\delta\hat{q}^m\delta\hat{q}^n-\langle
\delta\hat{q}^m\delta\hat{q}^n\rangle\right]p\nonumber\\
&-&\Gamma[f^\prime(q)]^2\delta\hat{p}+2f^\prime(q)\sum_{n\ge1}\frac{1}{n!}f^{n+1}(q)
\left[\delta\hat{q}^n\delta\hat{p}-\langle\delta\hat{q}^n\delta\hat{p}\rangle\right]
\nonumber\\
&+&\sum_{m\ge1}\sum_{n\ge1}\frac{1}{m!}\frac{1}{n!}f^{m+1}(q)f^{n+1}(q)
\left[\delta\hat{q}^m\delta\hat{q}^n\delta\hat{p}-\langle\delta\hat{q}^m\delta\hat{q}^n
\delta\hat{p}\rangle\right]\nonumber\\
&+&\eta(t)\left[f^{\prime\prime}(q)\delta\hat{q}+\sum_{n\ge 2
}\frac{1}{n!}f^{n+1}(q)[\delta\hat{q}^n-\langle\delta\hat{q}^n\rangle]\right]\label{A9}
\end{eqnarray}

The operator correction equations can be used to yield an infinite
hierarchy of equations.  Up to third order we construct, for
example, the following set of equations which are coupled to quantum
Langevin equations from (\ref{2.14}-\ref{2.15});
\begin{subequations}
\begin{eqnarray}
\frac{d}{dt}\langle\delta\hat{q}^2\rangle&=&\langle\delta\hat{q}\delta\hat{p}+
\delta\hat{p}\delta\hat{q}\rangle\label{A10a}\\
\frac{d}{dt}\langle\delta\hat{q}\delta\hat{p}+\delta\hat{p}\delta\hat{q}\rangle
&=&-2\chi(q,p)\langle\delta\hat{q}^2\rangle + 2
\langle\delta\hat{q}^2\rangle\nonumber\\
&-&\Gamma[f^\prime(q)]^2\langle\delta\hat{q}\delta\hat{p}+
\delta\hat{p}\delta\hat{q}\rangle
-\zeta(q,p)\langle\delta\hat{q}^3\rangle\nonumber\\
&-&2\Gamma
f^\prime(q)f^{\prime\prime}(q)\langle\delta\hat{q}^2\delta\hat{p}+
\delta\hat{p}\delta\hat{q}^2\rangle\label{A10b}\\
\frac{d}{dt}\langle\delta\hat{p}^2\rangle&=& - 2 \Gamma
[f^\prime(q)]^2\langle\delta\hat{p}^2\rangle-\chi(q,p)\langle\delta\hat{q}\delta\hat{p}+
\delta\hat{p}\delta\hat{q}\rangle\nonumber\\
&-&\frac{1}{2}\zeta(q,p)\langle\delta\hat{q}^2\delta\hat{p}+
\delta\hat{p}\delta\hat{q}^2\rangle\nonumber\\
&-&2\Gamma f^\prime(q)f^{\prime\prime}(q)
\langle\delta\hat{q}\delta\hat{p}^2+
\delta\hat{p}^2\delta\hat{q}\rangle\nonumber\\
\label{A10c}\\
\frac{d}{dt}\langle\delta\hat{q}^3\rangle&=&\frac{3}{2}\langle\delta\hat{q}^2\delta\hat{p}+
\delta\hat{p}\delta\hat{q}^2\rangle\label{A10d}\\
\frac{d}{dt}\langle\delta\hat{p}^3\rangle&=&-3 \Gamma
[f^\prime(q)]^2\langle\delta\hat{p}^3\rangle\nonumber
\\&-&\frac{3}{2}\chi(q,p)\langle\delta\hat{q}
\delta\hat{p}^2+\delta\hat{p}^2\delta\hat{q}\rangle\label{A10e}\\
\frac{d}{dt}\langle\delta\hat{q}^2\delta\hat{p}+
\delta\hat{p}\delta\hat{q}^2\rangle &=& -2
\chi(q,p)\langle\delta\hat{q}^3\rangle+2\langle\delta\hat{q}\delta\hat{p}^2+
\delta\hat{p}^2\delta\hat{q}\rangle\nonumber\\
&-&\Gamma[f^\prime(q)]^2\langle\delta\hat{q}^2\delta\hat{p}+
\delta\hat{p}\delta\hat{q}^2\rangle\label{A10f}\\
\frac{d}{dt}\langle\delta\hat{q}\delta\hat{p}^2+
\delta\hat{p}^2\delta\hat{q}\rangle &=& 2
\langle\delta\hat{p}^3\rangle-4\chi(q,p)\langle\delta\hat{q}^2\delta\hat{p}+
\delta\hat{p}\delta\hat{q}^2\rangle\nonumber\\
&-&2 \Gamma[f^\prime(q)]^2\langle\delta\hat{q}\delta\hat{p}^2
+\delta\hat{p}^2\delta\hat{q}\rangle\label{A10g}\\
where\;\;\;\;\;\;\;\;\;\;\;\;\;\;\;\;\;\;\;\;&&\nonumber\\
\chi(q,p)&=&V^{\prime\prime}(q)+2\Gamma\;p\;f^\prime(q)
f^{\prime\prime}(q)-\eta(t)f^{\prime\prime}(q)\nonumber
\\\label{A10h}\\
\zeta(q,p)&=&V^{\prime\prime\prime}(q)+2\Gamma\;p\;f^\prime(q)
f^{\prime\prime\prime}(q)
\nonumber\\&+&2\Gamma\;p\;[f^{\prime\prime}(q)]^2-
\eta(t)f^{\prime\prime\prime}(q)\label{A10i}
\end{eqnarray}
\end{subequations}

For other details we refer to \cite{barik}.

\end{appendix}


\begin{center}
{\bf Figure Captions}
\end{center}

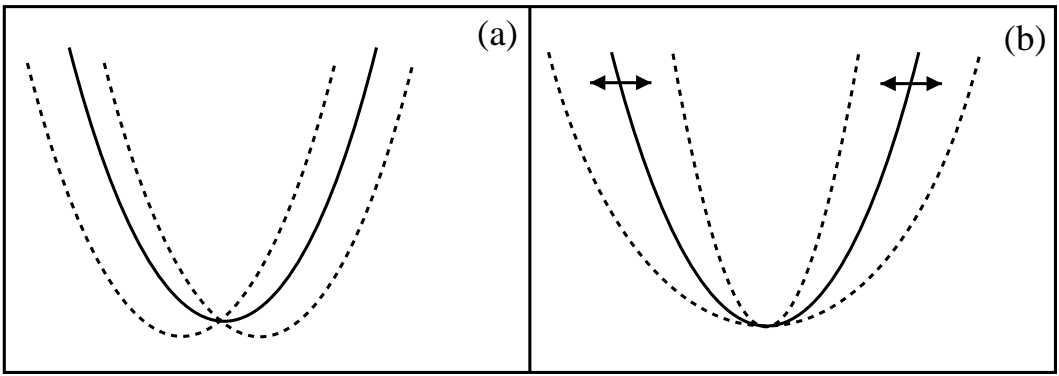
Fig.1: Schematic illustration of the effect of linear-linear(a) and
 square-linear(b) system-bath coupling on a relevant potential system.

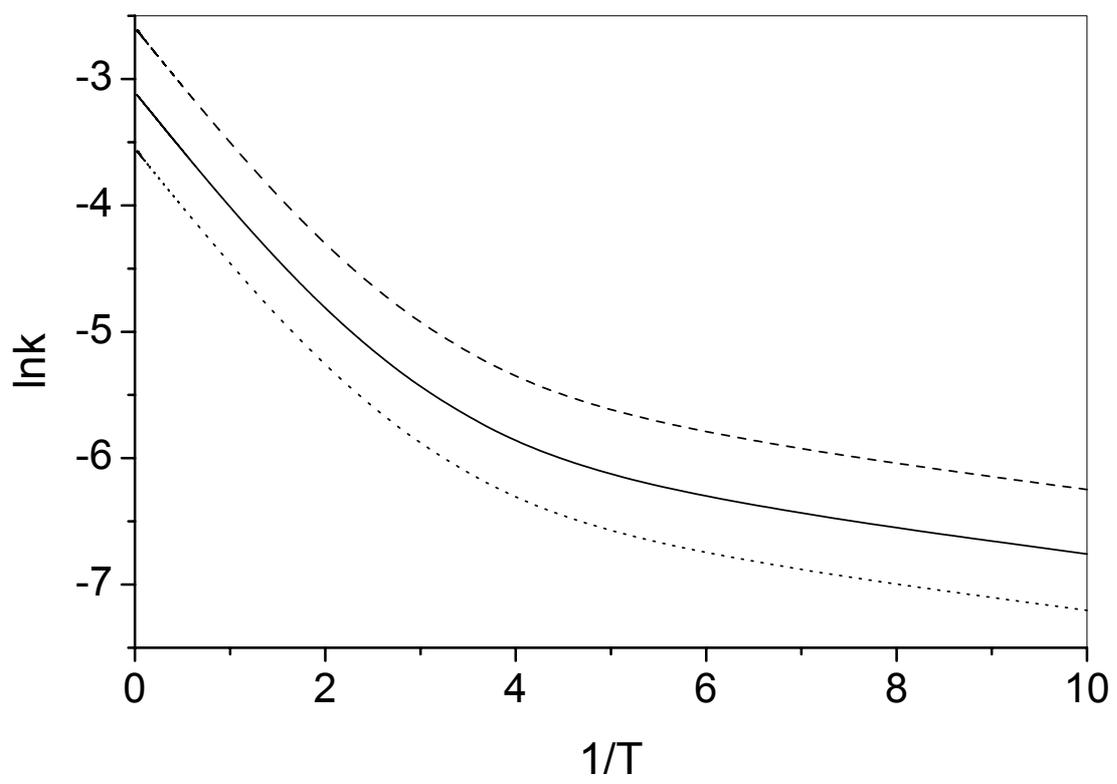
 Fig.2: Plot of  $\ln k$ vs. $1/T$ for different strength of
 nonlinear coupling $b_2$
((i)$b_2=0.01$ (dashed line) (ii) $b_2=0.1$ (solid line) (iii)
$b_2=0.2$ (dotted line) for a fixed parameter set
$a=0.5,\;b=0.15,\;b_1=1.0$ and $\Gamma=1.0$ (all the quantities are
in dimensionless unit).

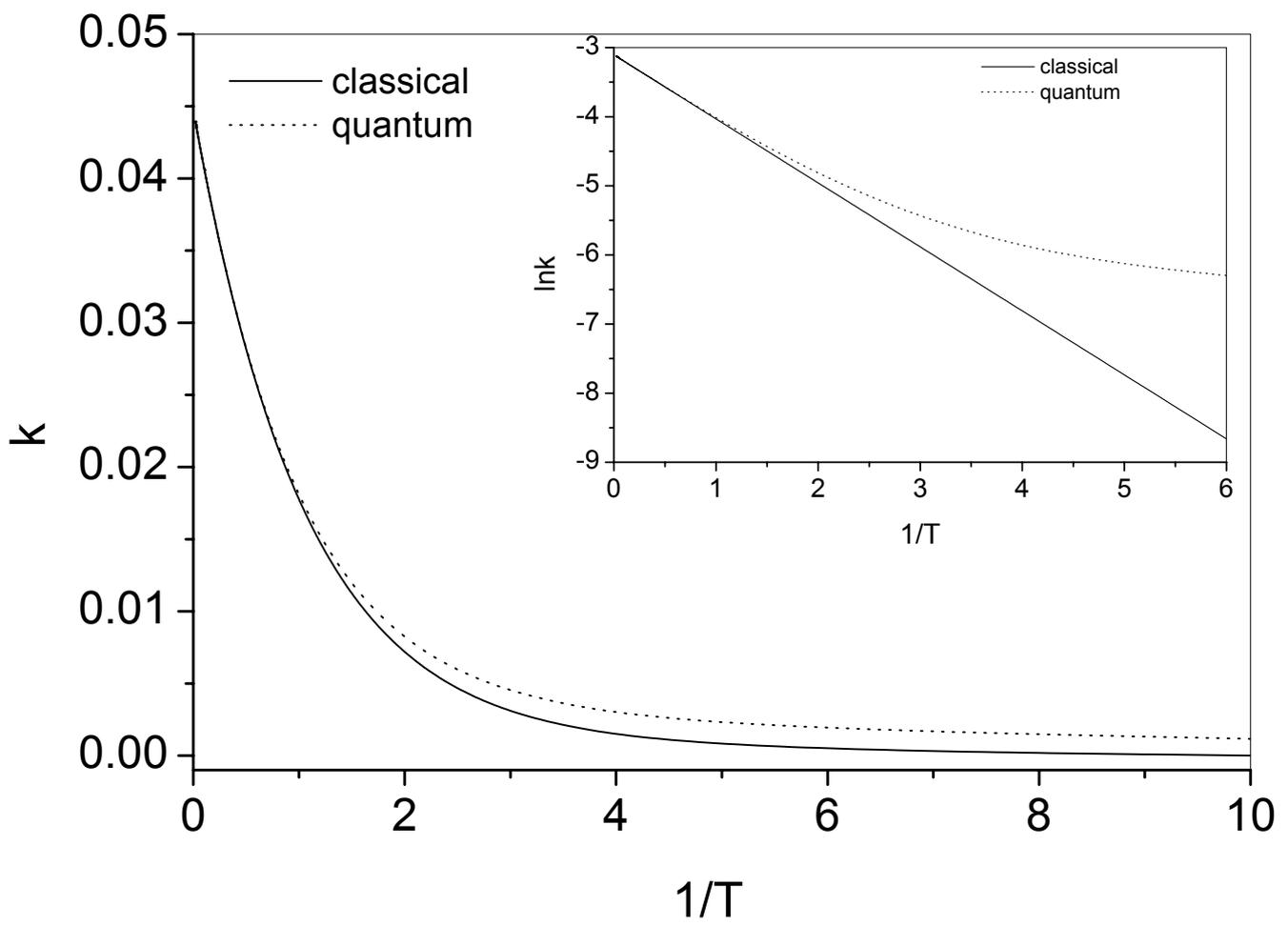
Fig.3: A comparison between classical(solid line) and quantum
(dotted line) decay rate plotting $k$ as a function of $1/T$ for the
parameter set $a=0.5,\;b=0.15,\;b_1=1.0,\;b_2=0.10$ and
$\Gamma=1.0$. The inset  presents a variation of $\ln k$  with $1/T$
comparing classical (solid line) and quantum (dotted line) cases for
the same parameter set as in the main figure.(all the quantities are
in dimensionless unit).


\begin{thebibliography}{200}
\bibitem{zwa} R. Zwanzig, Lectures in Theoretical Physics, Vol.3;
J. Stat. Phys. {\bf 9}, 215 (1973).
\bibitem{cald} A. O. Caldeira and A. J. Leggett, Ann. Phys. (N.Y) {\bf 149}, 374
(1983).
\bibitem{cald1} A. O. Caldeira and A. J. Leggett, Ann. Phys. (N.Y) {\bf 153},
445 (1984).
\bibitem{grab} H. Grabert, P. Schramm and G. L. Ingold, Phys. Rep. {\bf 168}, 115
(1988).

\bibitem{tani} K. Okumura and Y. Tanimura, Phys. Rev. E {\bf 56},
2747 (1997); T. Steffen and Y. Tanimura, J. Phys. Soc. Jap. {\bf
69}, 3115; T. Kato and Y. Tanimura, J. Chem. Phys. {\bf 117}, 6221
(2002); \textit{ibid.} {\bf 120}, 260 (2004).

\bibitem{db1} D. Barik, D. Banerjee and D. S. Ray, \textit{in Progress in Chemical Physics
Research}, Vol- 1, Edited by A. N. Linke, (Nova Publishers, New
York, ISBN: 1-59454-451-4, 2006).

\bibitem{wei} U. Weiss, \textit{Quantum Dissipative systems} (World Scientific,
Singapore, 1999).

\bibitem{san} J. M. Sancho, M. San Miguel, S. L. Katz and J. D.
Gunton, Phys. Rev. A {\bf 26}, 1589 (1982).

\bibitem{san1} J. M. Sancho, M. San Miguel and D. D\"{u}rr, J.
Stat. Phys. {\bf 28}, 291 (1982).

\bibitem{jay} A. M. Jayannavar and M. C. Mahato, Pramana {\bf 45},
368 (1995).

\bibitem{san2}A. Hernandez-Machado, M. SanMiguel and J. M. Sancho,
Phys. Rev. A {\bf 29}, 3388 (1984).

\bibitem{mas} J. Masoliver and L. Garrido, Phys. Lett. A {\bf
103}, 366 (1984)

\bibitem{lin1} J. D. Ramshaw and K. Lindenberg, J. Stat. Phys. {\bf
45}, 295 (1986).

\bibitem{sak} H. Sakaguchi, J. Phys. Soc. Jpn. {\bf 70}, 3247
(2001).

\bibitem{tai} C. Anteneodo and C. Tsallis, J. Math. Phys. {\bf
44}, 5194 (2003).

\bibitem{san3} J. Garcia-Ojalvo and J. M. Sancho, \textit{Noise in
spatially extended systems} (Springer-Verlag, New York, 1999).


\bibitem{barik} D. Barik and D. S. Ray, J. Stat. Phys. {\bf
120}, 339 (2005).

\bibitem{mill} W. H. Miller, S. D. Schwartz and J. W. Tromp, J. Chem. Phys. {\bf 79},
 4889 (1983); P. H\"{a}nggi, P. Talkner and M. Borkoves, Rev. Mod.
Phys., {\bf 62}, 251(1990).
\bibitem{stra} M. Bogu\"{n}\'{a}, J. M. Porr\.{a}, J. Masoliver and K. Lindenberg,
Phys. Rev. E {\bf 57}, 003990 (1998).

\bibitem{maka} D. E. Makarov and M. Topaler Phys. Rev. E {\bf 52},
178 (1995).

\bibitem{bao} J. D. Bao, Phys. Rev. A {\bf 65}, 052120
(2002); J. D. Bao, Phys. Rev. A {\bf 69}, 022102 (2004).

\bibitem{pk1} P. K. Ghosh, D. Barik, B. C. Bag and  D. S. Ray, J. Chem. Phys.
{\bf 123}, 224104 (2005).

\bibitem{but} M. B\"{u}ttiker, Z. Phys. B: Condensed Matter {\bf
68}, 161 (1987).

\bibitem{rei} P. Reimann, Phys. Rep. {\bf
361}, 57 (2002).

\bibitem{ff2} S. Faetti, P. Grigolini and F. Marchesoni,
Z. Phys. B {\bf 47}, 353 (1982).

\bibitem{ben} R. Benzi, G. Parisi, A. Sutera, and A. Vulpiani,
Tellus {\bf 34}, 10 (1982); B. McNamara, K. Wiesenfeld and R. Roy,
Phys. Rev. Lett. 60, 2626 (1988).

\bibitem{ff3} L. Gammaitoni, F.
Marchesoni, E. Menichella-Saetta and S. Santucci, Phys. Rev. Lett.
{\bf 62}, 349 (1989); L. Gammaitoni, F. Marchesoni, E. Menichella
-Saetta, and S. Santucci, Phys. Rev. E {\bf 49}, 4878 (1994); L.
Gammaitoni, P. H\"{a}nggi, P. Jung and F.Marchesoni, Rev. Mod.
Phys., {\bf 70}, 223(1998).

\bibitem{tran1} W. Horsthemke and R. Lefever, \textit{Noise-induced transitions:
Theory and applications in physics, chemistry, and biology}
(Springer-Verlag, Berlin and New York, 1984).

\bibitem{d41}P. K. Ghosh, D.
Barik and D. S. Ray, Phys. Lett. A {\bf 342} 12 (2005).
\bibitem{muka} S. Mukamel, \textit{Principles of Nonlinear Optical
Spectroscopy}(Oxford University Press, New York, 1995).

\bibitem{oxt} D. W. Oxtoby, Adv. Chem. Phys {\bf 40}, 1 (1979).

\bibitem{bader} J. S. Bader and B. J. Berne, J. Chem. Phys. {\bf 100},
8359 (1994).

\bibitem{polla} W. T. Pollard and R. A. Friesner, J. Chem. Phys. {\bf 100},
5054 (1994).

\bibitem{hil} E. P. Wigner, Phys. Rev. {\bf 40}, 749 (1932);
M. Hillery, R. F. O'Connell, M. O. Scully and E. P. Wigner, Phys.
Rep. {\bf 106}, 121 (1984).

\bibitem{lin} K. Lindenberg and V. Seshadri, Physica A, {\bf 109}, 483 (1981);
K. Lindenberg and E. Cort\'{e}s, \textit{ibid.} {\bf 126}, 489
(1984).

\bibitem{lui} W. H. Louisell, \textit{Quantum Statistical Properties of
Radiation} (J. Wiley, 1973).

\bibitem{skb} S. K. Banik, B. C. Bag and D. S. Ray,
Phys. Rev. E {\bf 65}, 051106 (2002).

\bibitem{db2}  P. K. Ghosh, D. Barik and D. S. Ray,
Phys. Rev. E {\bf 71}, 041107 (2005); D. Banerjee, B. C. Bag, S. K.
Banik and D. S. Ray, J. Chem. Phys. {\bf 120}, 8960 (2004)

\bibitem{bk1} D. Barik, S. K. Banik and D. S. Ray, J. Chem. Phys.
{\bf 119}, 680 (2003); D. Barik, B. C. Bag and D.S. Ray, J. Chem.
Phys. {\bf 119}, 12973 (2003).

\bibitem{bk2} D. Barik and D. S. Ray, J. Chem. Phys. {\bf 121}, 1681 (2004).

\bibitem{lan1} R. Landauer, Phys. Rev. A {\bf 12}, 636 (1975).

\bibitem{lan2} R. Landauer, J. Stat. Phys. {\bf 53}, 233 (1988).

\bibitem{mm1} P. H\"{a}nggi, F. Marchesoni and P. Grigolini, Z. Phys. {\bf B 56}, 333 (1984)
\bibitem{mm2} P. H\"{a}nggi, F. Marchesoni and P. Sodano,
 Phys. Rev. Lett. {\bf 60 }, 2563
(1988); P. H\"{a}nggi, F. Marchesoni and P. Riseborough Europhys.
Lett. {\bf 13}, 217 (1990).

\bibitem{mm5} J. B. Straus and G. A. Voth, J. Chem. Phys. {\bf 96},5460 (1991); J. B.
Straus, J. M. Liorente, and G. A. Voth, J. Chem. Phys. {\bf 98},4082
(1991); G. A. Voth, J. Chem. Phys. {\bf 97} 5908 (1993).
\end{thebibliography}
\end{document}